\pdfoutput=1
\documentclass[conference]{IEEEtran}
\IEEEoverridecommandlockouts

\usepackage{cite}
\usepackage{amsmath,amssymb,amsfonts}
\usepackage{algorithmic}
\usepackage{graphicx}
\usepackage{textcomp}
\usepackage{xcolor}
\usepackage{url}
\usepackage{eso-pic}
\AddToShipoutPictureBG*{%
  \AtPageLowerLeft{%
    \put(\LenToUnit{0.5\paperwidth},\LenToUnit{0.35in}){%
      \makebox[0pt][c]{\parbox{0.85\paperwidth}{\centering\scriptsize
        \copyright\,2026 IEEE. Personal use of this material is permitted. Permission from IEEE must be obtained for all other uses, in any current or future media, including reprinting/republishing this material for advertising or promotional purposes, creating new collective works, for resale or redistribution to servers or lists, or reuse of any copyrighted component of this work in other works. Accepted at the 2026 IEEE Spoken Language Technology Workshop (SLT).}}}}}

\def\BibTeX{{\rm B\kern-.05em{\sc i\kern-.025em b}\kern-.08em
    T\kern-.1667em\lower.7ex\hbox{E}\kern-.125emX}}
\begin{document}

\title{Controlling Speaking Rate in Autoregressive TTS via Activation Steering}

\author{\IEEEauthorblockN{Francesco Verdini$^{1,3,*}$, Antonis Asonitis$^{1,2,*}$, Aref Farhadipour$^{1,4}$, Marzieh Razavi$^{1}$,\\
Pierre-Edouard Honnet$^{1}$, Vijeta Avijeet$^{1}$, Juan Pablo Zuluaga Gomez$^{1}$}
\IEEEauthorblockA{$^{1}$AGIGO \quad $^{2}$ETH Zurich \quad $^{3}$Sapienza University of Rome \quad $^{4}$University of Zurich}
\thanks{$^{*}$These authors contributed equally.}
}

\maketitle

\begin{abstract}
Autoregressive text-to-speech (TTS) systems synthesize natural speech but, once trained, offer little control over speaking rate. We show that speaking rate can be steered at inference time, without retraining, by clamping a single decoder block's activation along a discovered speed axis. A decoder-block analysis recovers the rate axis, a neutral operating point, and a per-step intensity scale; at inference, the activation's projection onto this axis is set to a fixed scalar. Learning this direction from synthetically time-stretched and time-compressed speech yields rate control that largely preserves speaker identity, generalizes across model architectures, and maintains high naturalness in objective and human evaluations. Unlike standard additive steering, which breaks at the slow extreme, clamping remains stable on all three systems tested; at moderate targets, the better rule depends on the model. Finally, we show that rate information is decodable across layers but causally steerable only within a mid-depth window, and demonstrate the effectiveness of our approach on the public Seed-TTS-Eval benchmark.

\end{abstract}


\section{Introduction}
Modern autoregressive text-to-speech (TTS) systems synthesize remarkably natural speech \cite{wang2023valle,anastassiou2024seedtts,du2024cosyvoice2,hu2026qwen3tts,arora2025landscape,guo2025recent}, but offer little control over \emph{how} an utterance is delivered once the system is trained. Speaking rate is a basic case: text prompts~\cite{guo2023prompttts} and reference voices set it only coarsely, explicit duration control~\cite{ren2021fastspeech2} belongs to non-autoregressive models, and supervised fine-tuning for finer control is expensive or simply unavailable for an attribute the interface does not expose. In text language models a complementary, training-free approach has emerged: instead of changing inputs or weights, representation engineering and activation steering extract a direction in the residual stream for a target concept and intervene along it during generation \cite{zou2023repe,turner2023actadd,rimsky2023caa}. We ask whether the same idea can give natural, reliable control of speaking rate in autoregressive TTS, and find that it can, but only with a calibrated recipe (a rate-isolating contrast, a clamped application rule, and a causally chosen intervention layer) that preserves the high-quality voice cloning of these systems. Steering vectors have also reached audio diffusion models~\cite{staniszewski2026tada} and speech-aware LLMs~\cite{chang2026overcoming,lin2026steering,yegorova2026salsa}, e.g., to adapt ASR to out-of-domain speech~\cite{yegorova2026salsa}; we instead steer the generation side of a TTS decoder to control a perceptual attribute of its output.

\begin{figure}[t]
\centering
\includegraphics[width=\columnwidth]{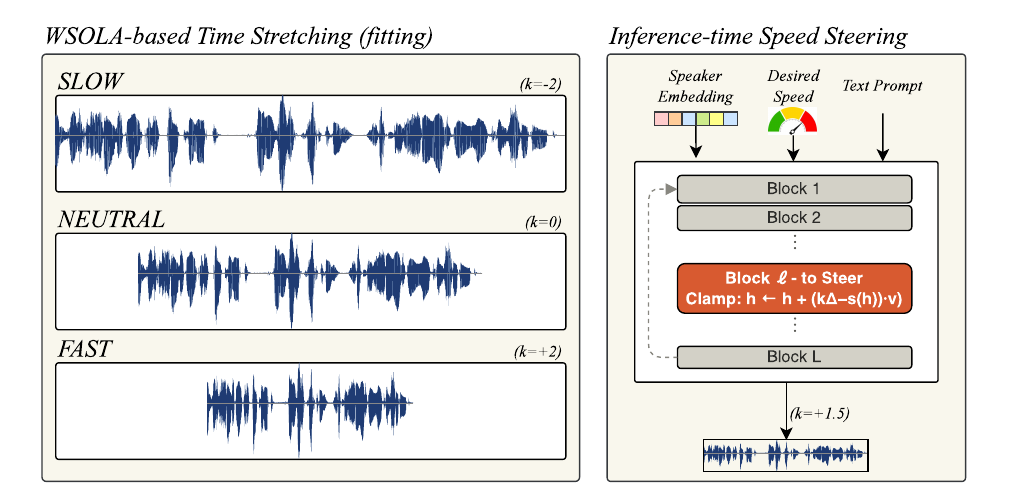}
\caption{Overview. Left: fitting. The rate axis is learned from WSOLA time-stretched speech, the same utterance at slow ($k{=}{-}2$), neutral ($k{=}0$), and fast ($k{=}{+}2$) tempo, teacher-forced through the frozen decoder. Right: inference. A single decoder block's activation is clamped along the learned axis to the requested intensity $k$, turning the pretrained model into a speaking-rate dial that moves tempo while preserving wording, voice, and naturalness. No fine-tuning.}
\label{fig:overview}
\end{figure}

The standard steering recipe is contrastive activation addition: estimate a steering vector from activation differences between positive and negative examples, then add that vector back at inference time \cite{subramani2022steering,turner2023actadd,rimsky2023caa,li2023iti}. The vector is simple to estimate, but the additive application rule is blunt: it applies the same fixed offset whether the current state is below, at, or already beyond the desired level, and its reliability is known to hinge on the geometry of the training contrast \cite{braun2026unreliable}. Prior work in text models already shows that useful interventions need not be purely additive: directional ablation zeros a projection onto a chosen direction \cite{arditi2024refusal}, affine concept editing resets an activation to a reference value \cite{marshall2024ace}, and sparse-autoencoder (SAE) feature clamping pins selected features to chosen values \cite{templeton2024scaling} or targets them with steering vectors~\cite{chalnev2024saets}. SAEs have also exposed interpretable features in language and audio models \cite{cunningham2024sae,mariotte2025saeaudio,aparin2026audiosae} and enabled emotion~\cite{du2026saetts} and speech-rate~\cite{koriagin2026saetts} control in LLM-based TTS, while activation steering has controlled emotion~\cite{wang2026cocoemo,zhou2026emoshift,xie2025emosteer} and accent~\cite{yang2026activation} in TTS and adapted audio-LLM ASR to accents~\cite{sun2026activation}. Unlike SAE-based rate steering~\cite{koriagin2026saetts}, we need no trained dictionary and clamp one probed direction to calibrated targets. We build on these state-setting operations and develop them into calibrated state-setting for controllable autoregressive TTS, moving one concrete, useful attribute, speaking rate, cleanly and reliably.

This \emph{how} matters most in autoregressive TTS, where each edited state is fed back as the history the model must continue while preserving the reference voice. A fixed additive offset has no state-dependent stopping condition: it keeps pushing in the same direction even when the trajectory is already at the requested rate, and over a long generation the pushes compound and drive the state out of the region the model was trained on. The symptom differs by system, over-fast continuations and non-termination, voice drift, or garbling, but the cause is shared. We therefore treat the application rule as part of the control method, not as an implementation detail.

Our recipe is to \emph{probe} and \emph{clamp}. From a small same-text slow/neutral/fast corpus teacher-forced through the system, we probe a decoder block to recover a rate axis, a neutral operating point, and an intensity unit. At inference we then \emph{set}, rather than add to, the activation's projection on that axis, clamping it to the requested value. This turns steering into feedback control: the correction is large when the state is far from target, vanishes at the target, and reverses sign after overshoot. Because a layer from which an attribute can be \emph{read} is not necessarily one at which it can be \emph{written}, we choose the intervention site with a small causal sweep rather than by probe accuracy. The method rests on three design choices: a rate-isolating contrast, a causally localized intervention layer, and a clamped application rule; the experiments test each in turn. Figure~\ref{fig:overview} sketches this setup.

We evaluate speaking-rate control in Qwen3-TTS \cite{hu2026qwen3tts}, MOSS-TTS \cite{gong2026mosstts}, and CosyVoice2 \cite{du2024cosyvoice2}, measuring not only how far the rate moves but what the movement costs in intelligibility, naturalness, voice preservation, and non-termination. Additive steering is adequate for short utterance edits but breaks as the target grows, while clamping holds a far wider stable range. The same calibrated intensities, tuned once on LibriTTS \cite{zen2019libritts} clean sentences, still move rate on a thousand unseen voices and texts in the public Seed-TTS-Eval benchmark \cite{anastassiou2024seedtts}.

In summary, we make the following contributions:
\begin{itemize}
\item We show that speaking rate in autoregressive TTS can be moved across a wide range at inference time, without retraining, while preserving intelligibility, naturalness, and voice identity; and that the application rule is decisive at strong intensities: additive edits compound over autoregressive generation and break at the slow extreme on all three systems, whereas clamping the projection prevents that accumulation; at moderate targets the better rule is model-dependent.
\item We show that a clean rate direction can be learned from synthetic data: time-stretching natural speech with the classical WSOLA algorithm \cite{verhelst1993wsola} changes tempo while preserving pitch and timbre, so the contrast isolates rate by construction and the recipe needs no separate disentanglement step.
\item We show that speaking rate is decodable across most of the decoder yet causally steerable only in a mid-depth window, so the intervention layer must be located by a causal sweep rather than assumed from probe accuracy.
\item We evaluate effect and side effects jointly on out-of-distribution LibriTTS speakers and confirm that the calibrated steering transfers to the public Seed-TTS-Eval benchmark, corroborated by a small human AB test.
\end{itemize}

\section{Method}\label{sec:method}

Our method for natural speaking-rate control combines three ingredients: we \emph{probe} the residual stream on a small same-text rate-contrast corpus to recover a calibrated rate axis (Sec.~\ref{ssec:probe}); we \emph{localize} the layer at which the axis is causally effective with a cheap sweep (Sec.~\ref{ssec:where}); and we \emph{steer} at inference by clamping the activation's projection onto the axis rather than adding to it (Sec.~\ref{ssec:clamp}).

\subsection{Setting}\label{ssec:setting}
We consider autoregressive TTS systems whose decoder generates discrete speech tokens~\cite{mousavi2025tokens,guo2025recent} (acoustic codec~\cite{zeghidour2021soundstream,defossez2022encodec,della2026focalcodec,della2026focalcodec_stream,siuzdak2024snac} or semantic) from a text prompt and a reference voice. Let $h \in \mathbb{R}^d$ denote the residual-stream output of decoder block $\ell$ at the current generation position. Every intervention below is a forward hook on block $\ell$ that rewrites only the hidden state of the last sequence position, i.e., the token currently being generated; model weights, prompts, and sampling parameters are otherwise unchanged.

\subsection{Probing the attribute axis}\label{ssec:probe}
Following TCAV~\cite{kim2018tcav}, we build the axis as a concept direction from a small same-text corpus in which the target attribute is recorded at three labelled values: a negative pole, a neutral center, and a positive pole. For speaking rate these are slow, neutral, and fast readings. Each recording is teacher-forced through the autoregressive TTS system, and we collect block-$\ell$ activations at speech-token positions. Because the texts are matched across classes, class-mean differences primarily reflect the target attribute rather than lexical content. From the class means $\mu_-,\mu_0,\mu_+$ we define
\begin{equation}
\hat v = \frac{\mu_+ - \mu_-}{\lVert \mu_+ - \mu_- \rVert},\qquad
c = \mu_0,\qquad
s(h) = \langle h - c,\, \hat v \rangle ,
\label{eq:axis}
\end{equation}
where $\hat v$ is the signed attribute direction, $c$ is the neutral operating point, and $s(h)$ is the activation's coordinate along the axis. The three means have distinct roles. The pole means $\mu_-$ and $\mu_+$ identify where the attribute changes and set the direction; the neutral mean $\mu_0$ fixes the origin of the command scale. Thus the neutral center is not an extra steering direction and does not make the intervention more fragile: it makes $k{=}0$ mean neutral speech.

We set the unit of the command scale from the labelled poles,
\begin{equation}
\Delta = \tfrac{1}{2}\bigl(\mathbb{E}_{+}[s(h)] - \mathbb{E}_{-}[s(h)]\bigr),
\label{eq:delta}
\end{equation}
where $\mathbb{E}_{\pm}$ averages over the speech-token activations $h$ of the fast/slow class, so the target $s^{\star}=k\Delta$ is expressed in class-gap units; $\Delta$ is thus the per-step intensity scale of the control, since at inference the intervention enforces the target $k\Delta$ at every generation step (Sec.~\ref{ssec:clamp}). Held-out logistic probes \cite{alain2016probes} confirm that the axis orders the labels monotonically at the class-mean level, which is the level at which steering sets a trajectory target. The neutral mean is consistently not the pole midpoint, so centering is a real calibration choice rather than a formality. With $\hat v$ and $\Delta$ fixed, changing $c$ only shifts the meaning of $k$ by a constant: a pole midpoint can reproduce the same clamp after an offset correction, whereas dropping the center entirely creates a model-specific shift of the operating point. 

\subsection{Steering by clamping, not adding}\label{ssec:clamp}
At every generation step, the standard steering rule adds the direction with a fixed coefficient,
\begin{equation}
h \;\leftarrow\; h + k\Delta\,\hat v \qquad \text{(additive)},
\label{eq:add}
\end{equation}
whereas the clamp \emph{sets} the projection to the target $s^{\star}=k\Delta$:
\begin{equation}
h \;\leftarrow\; h + \bigl(k\Delta - s(h)\bigr)\,\hat v \qquad \text{(clamp)}.
\label{eq:clamp}
\end{equation}
The difference is that of open-loop versus closed-loop control. The additive rule applies the same offset at every token regardless of where the activation already is; over an autoregressive trajectory the pushes compound (a voice-cloning decoder conditions on its own steered past as the voice it must continue, so each pushed token shifts the operating point for the next), and nothing stops the state from leaving the region the model was trained on. The clamp's perturbation $(k\Delta - s(h))\hat v$ instead adapts to the state: it shrinks to zero once generation already expresses the target level, its magnitude is bounded by the distance to the setpoint, and applying it twice is the same as applying it once. Viewed this way, the clamp generalizes interventions used on text LLMs: directional ablation \cite{arditi2024refusal} is a clamp to the fixed target $0$, and affine concept editing \cite{marshall2024ace} resets to a reference value before adding a fixed offset; probing supplies the graded, calibrated targets $k\Delta$ that turn the same operation into a dialable control. As in prior TTS steering~\cite{xie2025emosteer,wang2026cocoemo}, both rules have a variant that rescales the edited state to its pre-edit norm, $h \leftarrow h\,\lVert h_{\text{orig}}\rVert / \lVert h \rVert$, which removes the energy injected by the edit.

\subsection{Where to intervene}\label{ssec:where}
The intervention layer is selected causally, not by probe accuracy alone. A probe asks whether rate can be read from a layer; steering asks whether changing that coordinate at the layer changes the generated speech, echoing evidence that where knowledge is localized need not be where editing it works best \cite{hase2023localization}. Rate is decodable across much of the stack, but causal control appears only in a mid-depth window (Sec.~\ref{ssec:res-local}). We therefore run a small causal sweep over depth: for each candidate layer, temporary directions are built from a 200-clip-per-class subset, raw additive and clamp steering are applied at $\pm 2\Delta_\ell$ on a small held-out set, and the rate span is recorded. After the layer is chosen, the final steering direction is recomputed on the full activation bank at that site.

\section{Experimental Setup}\label{sec:setup}

\subsection{Models}
We evaluate on three open-weight autoregressive TTS systems spanning two model families and two architectures: \textbf{Qwen3-TTS-12Hz-0.6B-Base} \cite{hu2026qwen3tts}, whose autoregressive ``talker'' LM has 28 decoder blocks (we steer block 14); \textbf{MOSS-TTS} \cite{gong2026mosstts}, with a 36-block decoder (block 18); and \textbf{CosyVoice2-0.5B} \cite{du2024cosyvoice2}, a hybrid of a 24-block Qwen2.5-0.5B LM~\cite{qwen2025qwen25} and a flow-matching~\cite{lipman2023flow} acoustic model (we steer block 14 of the LM's residual stream, not the flow-matching stage). Each steering depth is the peak of the causal sweep of Sec.~\ref{ssec:where}, in every case close to $50\%$ relative depth.

\subsection{Attribute corpora}\label{ssec:corpora}
The rate axis is learned from \emph{synthetically time-stretched} natural speech, so that the three rate classes differ systematically mainly in tempo. From 500 LibriTTS \cite{zen2019libritts} dev-clean utterances ($2$--$12$\,s, $39$ speakers) we form, for each clip, a same-utterance slow / neutral / fast triplet: \emph{neutral} is the original recording, \emph{fast} is a WSOLA time-stretch to $1.4$--$2.0\times$ faster, and \emph{slow} a WSOLA stretch to $0.55$--$0.8\times$ slower (pitch-preserving overlap-add time-scaling, \texttt{ffmpeg atempo}~\cite{ffmpegatempo}). Because the three classes are the same utterance at three tempi, their class-mean activation difference isolates rate from pitch, timbre, lexical content, and vocal affect by construction, without a separate disentanglement step. We adopt this synthetic contrast after comparing it against two natural alternatives. The first alternative builds the axis the same way from \emph{real}, naturally rate-varying speech, binning the LibriTTS pool \emph{per speaker} into slow/neutral/fast by measured syllable rate. The resulting axis is also pitch-clean, but its slow and fast examples differ in lexical content, and this dilutes the class-mean contrast into a much weaker knob (Table~\ref{tab:realvssynth}). Its per-layer gap $\Delta_\ell$ is $6\times$ smaller, and it needs $3$--$6\times$ the coefficient to reach a target. The second alternative, an axis built from instruction-conditioned ``speak fast/slow'' speech, is often heavily entangled with pitch and affect, and correcting it by orthogonalizing against those axes introduces its own side effects on voice quality. Each clip is teacher-forced through each system at its steering layer, yielding roughly $20$--$32$k speech-token activations per class per model, from which we build the direction $\hat v$, center $c$, and unit $\Delta$ of Sec.~\ref{ssec:probe}. Throughout, the rate axis is this WSOLA direction. It could in principle encode overlap-add artifacts, but the natural rate-binned axis moves rate the same way, only more weakly, and clamp-steered audio beats an actual WSOLA stretch in UTMOS (5 of 6 cells) and, on Qwen3-TTS, in listener preference (Secs.~\ref{ssec:res-target}, \ref{ssec:res-abtest}).

\begin{table}[t]
\caption{Rate-axis source on Qwen3-TTS-0.6B (block 14, clamp): a direction learned from real rate-binned LibriTTS vs.\ from the synthetic WSOLA time-stretch. $\Delta_\ell$ is the per-layer class gap; UTMOS is no-reference naturalness; spk-sim is ECAPA cosine to the cloned voice.}
\label{tab:realvssynth}
\begin{center}
\begin{tabular}{lccc}
\hline
\textbf{Rate axis source} & $\Delta_\ell$ & \textbf{UTMOS} & \textbf{spk-sim} \\
\hline
Real (rate-binned LibriTTS) & $0.31$ & $3.11$ & $0.578$ \\
Synthetic (WSOLA)           & $\mathbf{1.85}$ & $\mathbf{3.36}$ & $\mathbf{0.600}$ \\
\hline
\end{tabular}
\end{center}
\end{table}

\subsection{Protocol and metrics}
Each (rule, intensity) cell synthesizes the same fixed 17-word sentence in unseen LibriTTS test-clean voices, held out from the dev-clean speakers used to build the direction, via reference-based voice cloning ($n{=}50$ voices per cell for Qwen3-TTS, $n{=}20$ for MOSS-TTS and CosyVoice2). Intensities are $k \in \{\pm1,\pm2,\pm3\}$, escalated to $\pm4,\pm6$ where a model tolerates $\pm3$ unscathed.
We report effect and side effects jointly: \emph{words per second} (wps) on silence-trimmed audio (the control axis); \emph{runaway rate}, the fraction of generations that hit the token cap without terminating; \emph{WER} (in \%) of an ASR transcript (Qwen3-ASR-0.6B \cite{qwenteam2026qwen3asr}) against the prompt text; \emph{UTMOS} naturalness (1--5) \cite{saeki2022utmos}; and \emph{speaker similarity}, the ECAPA-TDNN \cite{desplanques2020ecapa} cosine between the output and the reference voice.

Beyond the fixed-intensity sweep, which characterizes each model's tolerance and breaking behavior, we add a \emph{target-based} protocol (Sec.~\ref{ssec:res-target}) that measures efficiency and collateral damage at a requested effect size. Rather than fixing $k$, we fix the desired effect (a $\times1.25$ speed-up or a $\times0.75$ slow-down of the wps rate) and, per (model, rule), find the intensity $k^\star$ that reaches it: a calibration sweep over $k\in\{1,2,3\}$, escalated where a target is not yet crossed, with the first crossing interpolated in log-ratio space and unreached targets handled best-effort at the plateau. Each $k^\star$ is then evaluated on 20 held-out sentences (10 short, 10 long) $\times$ 5 voices ($n{=}100$ per cell).

\section{Results}\label{sec:results}

\subsection{Decodable everywhere, steerable in a window}\label{ssec:res-local}

\begin{figure}[t]
\centering
\includegraphics[width=\columnwidth]{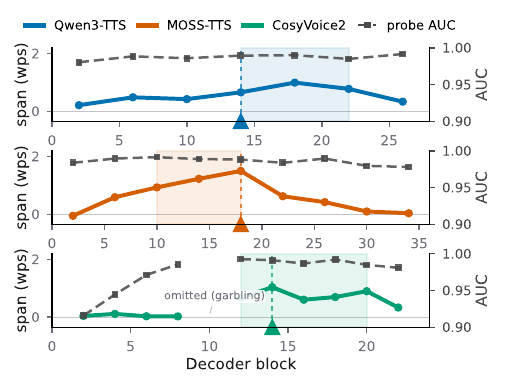}
\caption{Reading vs.\ writing speaking rate across depth. Solid: the wps span between generations clamp-steered at $k{=}{+}2$ and at $k{=}{-}2$ with the edit applied at that block alone (left axis; $n{=}10$ voices); a span near zero means intervening there does nothing. Dashed: held-out slow-vs-fast probe AUC at that block (right axis). Shading marks the causally steerable window; the triangle marks the selected steering block. AUC stays near ceiling across depth on all three systems while the span peaks mid-depth: rate is readable everywhere but writable only in a window.}
\label{fig:local}
\end{figure}

Figure~\ref{fig:local} asks two questions of every block of each system: whether rate can be \emph{read} there, via the AUC of a linear probe classifying slow vs.\ fast from that block's activations, and whether it can be \emph{written} there, via the span between generations steered at $k{=}{+}2$ and at $k{=}{-}2$ with the edit applied at that block alone. The two questions get opposite answers on all three systems. Reading succeeds essentially everywhere, with probe AUC near ceiling across depth, while writing works only in a mid-depth window: the span rises from essentially zero in the shallow blocks, is largest in a band around half depth that contains each system's selected steering site (Sec.~\ref{sec:setup}), and decays again toward the output. The implication is methodological: because probe accuracy is flat and saturated where the causal effect varies by an order of magnitude, it carries no information about where control is possible, so the intervention site must be located by a causal sweep rather than chosen by probe accuracy.

\subsection{The application rule decides at the extremes}\label{ssec:res-rule}

\begin{table}[t]
\caption{The slow extreme ($k=-2$) on all three systems: additive breaks, clamp holds. Baseline is the unsteered output. Run $=$ runaway rate (fraction of generations that never terminate); WER in \%; SIM $=$ ECAPA cosine to the cloned reference; \textbf{Bold} $=$ lower WER between the two rules for each model.}
\label{tab:cross}
\begin{center}
\begin{tabular}{llccccc}
\hline
 & \textbf{Rule} & wps & WER & UTMOS & SIM & Run \\
\hline
Qwen3-TTS  & Baseline  & 3.54 & 6.6 & 3.39 & 0.71 & 0\% \\
(block 14) & Additive  & 0.56 & 86.1 & 1.98 & 0.10 & 92\% \\
           & Clamp     & 2.14 & \textbf{8.6} & 2.76 & 0.52 & 0\% \\
\hline
MOSS-TTS   & Baseline  & 3.48 & 13.5 & 3.12 & 0.84 & 0\% \\
(block 18) & Additive  & 0.67 & 40.6 & 2.13 & 0.53 & 25\% \\
           & Clamp     & 2.05 & \textbf{9.1} & 2.93 & 0.79 & 0\% \\
\hline
CosyVoice2 & Baseline  & 3.44 & 7.3 & 3.52 & 0.81 & 0\% \\
(block 14) & Additive  & 2.13 & 55.3 & 2.98 & 0.73 & 0\% \\
           & Clamp     & 2.46 & \textbf{16.2} & 3.26 & 0.81 & 0\% \\
\hline
\end{tabular}
\end{center}
\end{table}

With the direction and intensity held fixed, only the application rule varies. Table~\ref{tab:cross} reports the slow extreme ($k{=}-2$) on all three systems; one intensity suffices because the axis separates additive from clamp at the same $k$ on every model. Additive breaks first, and worst on the slow side, each system in its own way: Qwen3-TTS barely slows while stripping the cloned voice and failing to terminate on nearly every generation, MOSS-TTS collapses in intelligibility and identity, and CosyVoice2 garbles without ever running away. Clamping the same projection at the same intensity holds all three, keeping error far below additive's, naturalness up, and speaker identity intact over a genuine slow-down. The failure signature differs by system, but in all three additive breaks first and clamp holds. Per-step renormalization of either rule leaves this picture unchanged, and the fast extreme is milder, so we report the slow side as the stress test; we use $k{=}-2$ because the synthetic axis's larger per-layer gap makes a step to $k{=}-3$ overshoot the slow pole, a property of the intensity scale rather than of the rule.

\begin{figure}[t]
\centering
\includegraphics[width=\columnwidth]{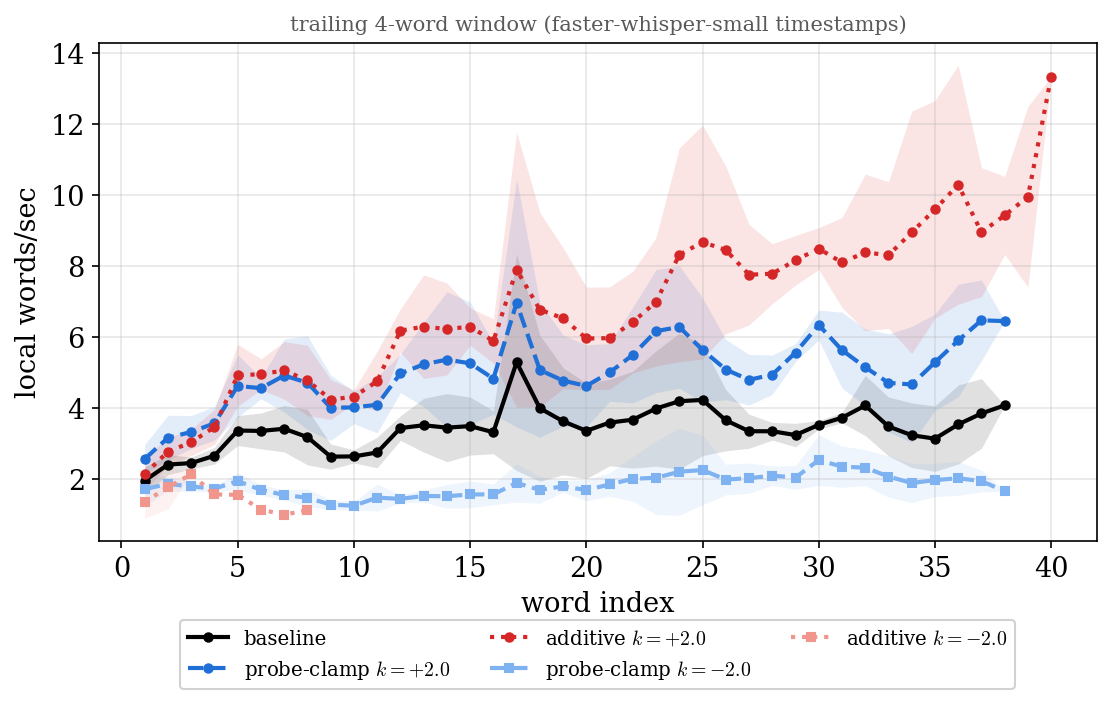}
\caption{Additive drift unfolding in time on Qwen3-TTS: local speaking rate (words per second in a trailing 4-word window, word timestamps from faster-whisper-small~\cite{fasterwhisper}) versus word index over a single very long sentence constructed for this analysis, averaged over 10 voices synthesizing the same sentence; bands are $\pm 1$ standard deviation. The additive offset compounds over generation, so additive $k{=}{+}2$ keeps accelerating along the utterance while the clamp at the same intensity settles early and holds a steady rate. The additive $k{=}{-}2$ line is interrupted where the model starts failing: beyond that point it does not say the rest of the sentence.}
\label{fig:longdrift}
\end{figure}

Figure~\ref{fig:longdrift} shows this breaking behavior unfolding in time on Qwen3-TTS, tracking the local rate word by word over one very long sentence: additive $k{=}{+}2$ never levels off, its rate climbing throughout the utterance as the fixed offset compounds, whereas the clamp at the same intensity reaches its faster rate within a few words and holds it. On the slow side the contrast is starker still: additive $k{=}{-}2$ fails within the first few words and never says the rest of the sentence, while the clamp sustains a genuine slow-down to the end of the text.

\subsection{Quality at practical rate targets}\label{ssec:res-target}

\begin{table*}[t]
\caption{Quality at matched moderate rate targets ($\times0.75$ slow / $\times1.25$ fast), two rules per model. $k^\star$ = the intensity magnitude that reaches the target (slow is negative); ach. = achieved wps ratio vs.\ baseline; WER/UTMOS/SIM at $k^\star$. \textbf{Bold} = lower WER between the two rules for a given model and direction. $^\dagger$target not reached (best-effort plateau), so quality is measured at a milder achieved rate. WSOLA = a time-stretch of the unsteered output to the target rate (DSP baseline); the rate matches the target by construction, so $k^\star$ is not applicable.}
\label{tab:target}
\begin{center}
\begin{tabular}{ll|ccccc|ccccc}
\hline
 & & \multicolumn{5}{c|}{\textbf{Slow target} ($\times0.75$)} & \multicolumn{5}{c}{\textbf{Fast target} ($\times1.25$)} \\
\textbf{Model} & \textbf{Rule} & $k^\star$ & ach. & WER (\%) & UTMOS & SIM & $k^\star$ & ach. & WER (\%) & UTMOS & SIM \\
\hline
Qwen3-TTS  & Additive  & 0.62 & 0.85$^\dagger$ & \textbf{0.6} & 3.34 & 0.61 & 0.79 & 1.30 & \textbf{0.7} & 3.24 & 0.54 \\
(block 14) & Clamp     & 1.37 & 0.74 & 0.7 & 3.23 & 0.58 & 1.16 & 1.25 & \textbf{0.7} & 3.28 & 0.56 \\
           & WSOLA     & -- & 0.75 & 0.8 & 2.85 & 0.61 & -- & 1.25 & 0.7 & 3.14 & 0.57 \\
\hline
MOSS-TTS   & Additive  & 0.66 & 0.79 & \textbf{5.0} & 3.04 & 0.79 & 0.99 & 1.32 & 11.2 & 2.83 & 0.77 \\
(block 18) & Clamp     & 1.47 & 0.72 & 8.5 & 2.91 & 0.78 & 0.66 & 1.24 & \textbf{5.9} & 2.95 & 0.78 \\
           & WSOLA     & -- & 0.75 & 10.4 & 2.60 & 0.78 & -- & 1.25 & 10.4 & 2.96 & 0.77 \\
\hline
CosyVoice2 & Additive  & 1.22 & 0.72 & \textbf{1.1} & 3.36 & 0.79 & 1.39 & 1.16$^\dagger$ & \textbf{4.1} & 3.44 & 0.77 \\
(block 14) & Clamp     & 1.73 & 0.73 & 1.7 & 3.37 & 0.79 & 1.66 & 1.23 & 17.6 & 3.40 & 0.76 \\
           & WSOLA     & -- & 0.75 & 1.1 & 2.83 & 0.77 & -- & 1.25 & 0.7 & 3.19 & 0.74 \\
\hline
\end{tabular}
\end{center}
\end{table*}

Table~\ref{tab:cross} compares the rules at their breaking point; Table~\ref{tab:target} re-asks the question at the rates a user actually requests, a $\times1.25$ speed-up and a $\times0.75$ slow-down. Here the picture is gentle: the requested effects need only small intensities, and on Qwen3-TTS and MOSS-TTS both rules stay essentially at baseline with no runaway, so the non-termination that defines Qwen3-TTS's slow extreme is an over-steering artifact that disappears at a practical target.

CosyVoice2 has a slightly different behavior: it garbles after steering, so even a moderate fast clamp raises its error and additive is the clean route there. The caveat for additive is temporal: because it applies the same offset at every step, it cannot back off once the history is already at rate, so the rate drifts over the course of a long utterance, and an utterance-level average hides this drift on a short prompt. The clamp's closed-loop correction matters most at the extremes and on long utterances; at moderate targets the simplest rule that stays within the model's tolerance is usually enough, with the clamp the safe default when length or intensity grows. Against the WSOLA stretch, the clamp has higher UTMOS in 5 of 6 cells (MOSS-TTS fast ties) and similar SIM, but higher WER on CosyVoice2 ($1.7$/$17.6\%$ vs.\ $1.1$/$0.7\%$, slow/fast).

\subsection{Generalization to a public benchmark (Seed-TTS-Eval)}\label{ssec:res-seedtts}
To test whether the calibration holds beyond our held-out sentences, we run on the full English Seed-TTS-Eval test set \cite{anastassiou2024seedtts,seedttseval}: the 1088 released Common Voice~\cite{ardila2020common} utterances over 666 prompt voices, each cloning its own prompt and synthesizing its own target text. We apply additive and clamp at the per-model fast/slow intensities calibrated once for the $\times1.25$/$\times0.75$ targets, with \emph{no} re-tuning for these voices. So that the numbers follow the public protocol, we adopt Seed-TTS-Eval's own evaluators: Whisper-large-v3 \cite{radford2023whisper} for WER, and its released WavLM-Large \cite{chen2022wavlm} speaker-verification checkpoint with an ECAPA-TDNN \cite{desplanques2020ecapa} pooling head for SIM.

The approach transfers (Table~\ref{tab:seedtts}): on all three systems the rate moves to $\times0.68$--$0.77$ for the requested $\times0.75$ and to $\times1.21$--$1.33$ for the requested $\times1.25$ across a thousand unseen voices and texts, at intensities well below what the raw natural rate-binned axis requires (Sec.~\ref{ssec:corpora}). On Qwen3-TTS and MOSS-TTS the move is inexpensive, with WER staying close to baseline. CosyVoice2 again shows its garbling signature: additive stays near baseline while the fast clamp garbles, the same failure mode as the controlled study, now on a public benchmark. As in Sec.~\ref{ssec:res-target}, the rule is chosen within each model's tolerance, and for CosyVoice2's fast direction that rule is additive. Speaker similarity is largely preserved at these practical targets: across the 1088 utterances, SIM stays within $0.11$ of each model's baseline, with the worst case at Qwen3-TTS clamp fast ($0.581 \to 0.471$). To anchor what that movement means, we score all $666\times665=442{,}890$ ordered different-speaker pairs within the benchmark's own prompt-voice pool under the same checkpoint. The resulting impostor similarities have a median of $0.087$ and a 99.9th percentile of $0.466$, with the single hardest pair reaching $0.677$. Every steered condition above, including the worst case, sits above that 99.9th percentile: each steered output remains closer to its true speaker than 999 of every 1000 random different-speaker pairs in the benchmark. 

\begin{table}[t]
\caption{Steering on the English Seed-TTS-Eval test set (1088 utterances, each cloning its own prompt and synthesizing its own target text). Additive and clamp applied at the per-model matched-effect intensities. $\times$base $=$ achieved wps ratio to the unsteered baseline; WER via Whisper-large-v3; SIM via Seed-TTS-Eval's own speaker-verification checkpoint (WavLM-Large + ECAPA-TDNN, \texttt{wavlm\_large\_finetune.pth}). \textbf{Bold} $=$ lower WER between the two rules for a given system and direction (both if tied).}
\label{tab:seedtts}
\begin{center}
\resizebox{\columnwidth}{!}{%
\begin{tabular}{ll|ccccc}
\hline
\textbf{System} & \textbf{Rule} & $k$ & wps & $\times$base & WER (\%) & SIM \\
\hline
Qwen3-TTS  & baseline      & ---       & 2.83 & 1.00 & 1.4 & 0.581 \\
           & additive fast & $+0.79$   & 3.45 & 1.22 & \textbf{1.4} & 0.514 \\
           & additive slow & $-0.62$   & 2.16 & 0.77 & \textbf{2.4} & 0.535 \\
           & clamp fast    & $+1.16$   & 3.55 & 1.26 & 1.6 & 0.471 \\
           & clamp slow    & $-1.37$   & 1.93 & 0.68 & 2.7 & 0.526 \\
\hline
MOSS-TTS   & baseline      & ---       & 2.82 & 1.00 & 7.0 & 0.740 \\
           & additive fast & $+0.99$   & 3.74 & 1.33 & \textbf{7.5} & 0.701 \\
           & additive slow & $-0.66$   & 2.15 & 0.76 & 5.3 & 0.729 \\
           & clamp fast    & $+0.66$   & 3.68 & 1.30 & 7.9 & 0.700 \\
           & clamp slow    & $-1.47$   & 2.08 & 0.74 & \textbf{5.0} & 0.728 \\
\hline
CosyVoice2 & baseline      & ---       & 2.72 & 1.00 & 2.4 & 0.659 \\
           & additive fast & $+1.39$   & 3.29 & 1.21 & \textbf{4.2} & 0.641 \\
           & additive slow & $-1.22$   & 2.03 & 0.74 & \textbf{3.0} & 0.660 \\
           & clamp fast    & $+1.66$   & 3.52 & 1.29 & 17.6 & 0.622 \\
           & clamp slow    & $-1.73$   & 2.04 & 0.75 & \textbf{3.0} & 0.660 \\
\hline
\end{tabular}%
}
\end{center}
\end{table}

\subsection{Human evaluation: competitive with a time-stretch baseline}\label{ssec:res-abtest}
Steering the rate inside the model is useful in its own right: a time-stretch is a second pass over finished audio, whereas the clamp is a per-step decoder hook that emits audio at the target rate directly, a natural fit for streaming. It also provides the human check on naturalness: listeners compare clamp-steered audio against a matched-rate WSOLA time-stretch of the same unsteered utterance (same wording, same tempo), so the forced choice isolates perceived quality, with the time-stretch, the standard post-hoc alternative, as the reference. Across 800 comparisons (36 listeners in total: 20 per direction, 20 pairs each, with four listening to both directions; one further listener, who completed only 18 of 20 pairs, was excluded), listeners preferred the steered audio in 48.2\% of cases and heard no difference in 11.0\%, against 40.8\% for the time-stretch; the pooled preference is significant (two-sided exact sign test on decisive votes, $p{=}0.027$), robust to listener clustering (listener-level bootstrap over the 36 listeners, 10{,}000 resamples: clamp share of decisive votes $0.542$, 95\% CI $[0.507, 0.575]$), and directionally consistent in both the slow and fast conditions, though not individually significant in either (Table~\ref{tab:abtest}). We read this small study as finding clamp steering at least competitive with, and overall slightly preferred over, matched-rate WSOLA: steering at calibrated intensities does not cost perceived quality relative to the standard alternative, corroborating the objective naturalness metrics.

\begin{table}[t]
\caption{Forced-choice preference: clamp steering vs.\ a matched-rate time-stretch of the unsteered output (Qwen3-TTS; 36 listeners: 20 per direction, 20 pairs each). Cells are the share of comparisons (clamp / tie / time-stretch); $p$ is a two-sided exact sign test on decisive votes.}
\label{tab:abtest}
\begin{center}
\begin{tabular}{lcccc}
\hline
\textbf{Condition} & \textbf{Clamp} & \textbf{Tie} & \textbf{Time-stretch} & \textbf{$p$} \\
\hline
Slow     & \textbf{48.2\%} & 10.5\% & 41.2\% & 0.153 \\
Fast     & \textbf{48.2\%} & 11.5\% & 40.2\% & 0.099 \\
Combined & \textbf{48.2\%} & 11.0\% & 40.8\% & 0.027 \\
\hline
\end{tabular}
\end{center}
\end{table}

\section{Limitations}\label{sec:limitations}

The clean, single-attribute rate axis that anchors our method comes for free only because speaking rate has a faithful synthetic transform: a time-stretch that changes tempo while preserving pitch and timbre. Attributes that lack such an attribute-isolating transform, for instance pitch, timbre, or emotion, would not yield a clean contrast this way and would fall back on instructed or natural data with its entanglement, so how far the recipe extends beyond speaking rate remains open. The study is also scoped in other respects: we steer one attribute at the granularity of a whole utterance and do not show time-varying control within a sentence or a streaming deployment, we evaluate in English only (LibriTTS and Seed-TTS-Eval), and we rely largely on objective proxies for word error rate, naturalness, and speaker similarity, which capture intelligibility, quality, and identity only approximately. Finally, the axis is one linear direction from a single stretching algorithm; we do not test nonlinear rate geometry, the axis's stability across data size and speakers, or the per-utterance spread of the achieved rate, and $k$ is calibrated per model and rule rather than predicted.

\section{Conclusion}\label{sec:conclusion}
We have shown that speaking rate in autoregressive TTS can be controlled at inference time, without retraining, by clamping one decoder block's activation along a rate axis learned from WSOLA time-stretched speech, at a block located by a causal sweep rather than by probe accuracy. Across Qwen3-TTS, MOSS-TTS, and CosyVoice2, additive steering breaks first at the extremes while clamping holds a far wider stable range; at moderate targets the gentler additive rule is often sufficient, and on CosyVoice2's fast side it is the cleaner route, so the clamp is the usual but not universal safe choice. Calibrated once, the same intensities move rate on a thousand unseen Seed-TTS-Eval voices, and listeners slightly prefer the clamped output over a matched-rate time-stretch.

\section*{Acknowledgment}
Claude Code (Anthropic; Claude Opus 4.7--5) and local open-weight LLMs (Qwen, DeepSeek, Kimi, GLM) assisted with the experimental code; these and ChatGPT (OpenAI; GPT-5.5/5.6) assisted with grammar and wording in all sections. All outputs were reviewed by the authors.

\bibliographystyle{IEEEtran}
\bibliography{references}

\end{document}